\documentclass[aps,prl,reprint,superscriptaddress]{revtex4-2}

\usepackage{stackengine,graphicx,xcolor}
\usepackage{siunitx}
\usepackage{physics}
\AtBeginDocument{\RenewCommandCopy\qty\SI}
\usepackage{amsmath}
\usepackage{etaremune}
\usepackage{braket}
\usepackage{mathtools}
\usepackage{anyfontsize}
\usepackage{appendix}
\usepackage{multirow}
\usepackage[hidelinks]{hyperref}
\usepackage[document]{ragged2e}
\usepackage[export]{adjustbox}
\usepackage{footnote}
\usepackage{lipsum}
\makesavenoteenv{table}
\usepackage[version=4]{mhchem}
\usepackage{nicefrac}
\usepackage{orcidlink}
\usepackage{svg}

\usepackage[sectionbib]{bibunits}

\defaultbibliographystyle{apsrev4-2} 
\defaultbibliography{references_wg}

\usepackage{hyperref}
\hypersetup{
    colorlinks=true,
    linkcolor=blue,
    urlcolor=magenta,      
    citecolor=cyan,
    }

\usepackage{float}
\restylefloat{table}

\newcommand{\refsub}[2]{\hyperref[#1]{\ref*{#1}(#2)}}
\newcommand{\s}[1]{\mathrm{#1}}

\makeatletter
\def\maketitle{
\@author@finish
\title@column\titleblock@produce
\suppressfloats[t]}
\makeatother

\begin{document}

\title{
Nonlinear Quantum Photonics with a Tin-Vacancy Center Coupled to a One-Dimensional Diamond Waveguide}

\author{Matteo Pasini\orcidlink{0009-0005-1358-7896}}
\affiliation{QuTech and Kavli Institute of Nanoscience, Delft University of Technology, PO Box 5046, 2600 GA Delft, The Netherlands}

\author{Nina Codreanu\orcidlink{0009-0006-6646-8396}}
\affiliation{QuTech and Kavli Institute of Nanoscience, Delft University of Technology, PO Box 5046, 2600 GA Delft, The Netherlands}

\author{Tim Turan\orcidlink{0009-0003-9908-7985}}
\affiliation{QuTech and Kavli Institute of Nanoscience, Delft University of Technology, PO Box 5046, 2600 GA Delft, The Netherlands}

\author{Adri\`a~Riera~Moral}
\thanks{Present address: Sparrow Quantum, Copenhagen, Denmark}
\affiliation{QuTech and Kavli Institute of Nanoscience, Delft University of Technology, PO Box 5046, 2600 GA Delft, The Netherlands}

\author{Christian~F.~Primavera\orcidlink{0000-0001-7486-6243}}
\affiliation{QuTech and Kavli Institute of Nanoscience, Delft University of Technology, PO Box 5046, 2600 GA Delft, The Netherlands}

\author{Lorenzo~De~Santis\orcidlink{0000-0003-0179-3412}}
\affiliation{QuTech and Kavli Institute of Nanoscience, Delft University of Technology, PO Box 5046, 2600 GA Delft, The Netherlands}

\author{Hans~K.~C.~Beukers\orcidlink{0000-0001-9934-1099}}
\affiliation{QuTech and Kavli Institute of Nanoscience, Delft University of Technology, PO Box 5046, 2600 GA Delft, The Netherlands}

\author{Julia~M.~Brevoord\orcidlink{0000-0002-8801-9616}}
\affiliation{QuTech and Kavli Institute of Nanoscience, Delft University of Technology, PO Box 5046, 2600 GA Delft, The Netherlands}

\author{Christopher Waas\orcidlink{0009-0008-1878-2051}}
\affiliation{QuTech and Kavli Institute of Nanoscience, Delft University of Technology, PO Box 5046, 2600 GA Delft, The Netherlands}

\author{Johannes Borregaard}
\thanks{Present address: Department of Physics, Harvard University, Cambridge, Massachusetts 02138, USA}
\affiliation{QuTech and Kavli Institute of Nanoscience, Delft University of Technology, PO Box 5046, 2600 GA Delft, The Netherlands}

\author{Ronald Hanson\orcidlink{0000-0001-8938-2137}}
\email{r.hanson@tudelft.nl}
\affiliation{QuTech and Kavli Institute of Nanoscience, Delft University of Technology, PO Box 5046, 2600 GA Delft, The Netherlands}

\begin{abstract}
    \justifying

    \noindent Color-centers integrated with nanophotonic devices have emerged as a compelling platform for quantum science and technology.
    Here we integrate tin-vacancy centers in a diamond waveguide and investigate the interaction with light at the single-photon level. We observe single-emitter induced extinction of the transmitted light up to $25\%$ and measure the nonlinear effect on the photon statistics. Furthermore, we demonstrate fully tunable interference between the reflected single-photon field and laser light back-scattered at the fiber end and show the corresponding controlled change between bunched and anti-bunched photon statistics in the reflected field.

\end{abstract}

\maketitle

\begin{bibunit}

\justifying
Nonlinear interactions between single photons and solid-state color centers are at the heart of many applications in quantum science \cite{awschalom_quantum_2018, atature_material_2018} such as the realization of a quantum internet \cite{kimble_quantum_2008, wehner_quantum_2018}. 
In particular, color centers in diamond have enabled advanced demonstrations in this direction showing multinode quantum network operation 
\cite{pompili_realization_2021, hermans_qubit_2022}, memory-enhanced communication
\cite{bhaskar_experimental_2020}
and scalable on-chip hybrid integration \cite{wan_large-scale_2020}. 
Among the diamond color centers, the tin-vacancy center (SnV) has recently emerged as a promising qubit platform as it combines the inversion symmetry of group-IV color centers~\cite{bradac_quantum_2019, ruf_quantum_2021}, allowing for integration in nanophotonic structures, with good optical properties~\cite{trusheim_transform-limited_2020, arjona_martinez_photonic_2022, rugar_quantum_2021,brevoord_heralded_2023} and above-millisecond spin coherence at temperatures above 1~K~\cite{rosenthal_microwave_2023, guo_microwave-based_2023}. Devices combining photonic integration with spin and optical control could serve as a future scalable building block for realizing spin-photon gates~\cite{beukers_tutorial_2023}.
On the path towards such scalable on-chip integration, incorporation of emitters into nanophotonic waveguides \cite{sipahigil_integrated_2016, arjona_martinez_photonic_2022} enables exploration of the emitter-photon interaction and probing of the resulting optical signals both in reflection and transmission.

\begin{figure*}[t]
    \centering
    \includegraphics[width=\textwidth]{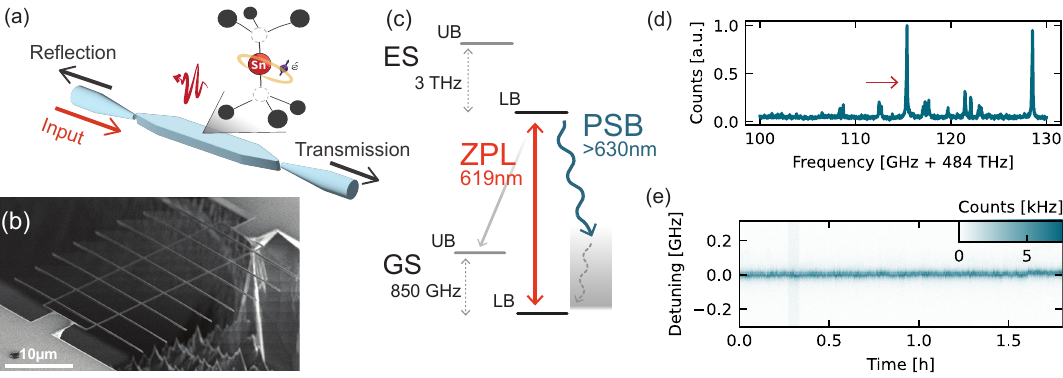}
    \caption{Device and optical transition properties. 
    (a) Schematic of the device. We address SnV centers in a nanophotonic diamond waveguide with two tapered ends, to which we couple using tapered optical fibers.
    (b) SEM image of one entire diamond device chip.
    (c) Energy level scheme of the SnV. At \qty{5}{\K} the SnV spectrum shows two ZPL transitions between the Lower Branch (LB) of the Excited State (ES) to the Lower (Upper) branches (LB, UB) of the Ground State (GS). Here we focus on the ZPL transition between LB of ES and LB of GS at \qty{619}{nm}, and we filter the other ZPL out. Phonon assisted decay from the excited state gives rise to a phonon sideband (PSB) with a broad optical spectrum above \qty{630}{\nm}.
    (d) Photoluminescence excitation in one waveguide showing several SnVs. The red arrow indicates the SnV used for this experiment.
    (e) Consecutive PLE scans conditioned on the SnV being in the right charge and frequency state. The scanning speed of each PLE is $\sim \qty{300}{\MHz/\second }$.}
    \label{fig:fig1}
\end{figure*}

In this work, we present a device consisting of a SnV center coupled to a nanophotonic diamond waveguide with tapered-fibre access on both sides, depicted in Fig.~\refsub{fig:fig1}{a}, and we show its nonlinear interaction at the single-photon level with a weak coherent laser field. 
We observe the extinction of transmitted light which arises from interference between the photons interacting with the emitter and the single optical mode of the waveguide~\cite{chang_single-photon_2007, shen_coherent_2005}, from which we quantify the coupling strength of the SnV to the waveguide. 
In the reflected signal we observe interference between single photons scattered by the emitter and a classical reflection of the probe laser.
By measuring photon correlations we observe that the emitter interacts with the incident field at the single photon level, altering its photon statistics both in the transmitted and in the reflected signals. 
The efficient coupling between SnV and waveguide, together with the single photon nature of the interaction, make this system a promising starting point for quantum photonic applications.

Our device is fabricated in two main phases: the generation of SnV centers in the bulk diamond substrate, followed by the nanofabrication of suspended waveguides that support a single TE mode for SnV emission.
An electronic grade diamond substrate is pre-processed and implanted with \ce{^{120}Sn} ions at a target depth of $\sim\qty{88}{\nm}$, followed by an annealing step to create SnV centers.
The nanofabrication of the waveguides is based on the crystal-dependent quasi-isotropic-etch undercut method \cite{khanaliloo_high-qv_2015, mitchell2019realizing, mouradian_rectangular_2017, wan_large-scale_2020, rugar_quantum_2021, ruf_cavity-enhanced_2021}. 
To fabricate the waveguide chips, we first pattern a hard mask material \ce{Si3N4}, followed by the transfer pattern into the diamond substrate and vertical coverage with \ce{Al2O3} of the structures sidewalls.
Next, the quasi-isotropic etch undercuts the devices, followed by an upward etch to thin the devices down to a thickness of $\approx \qty{250}{nm}$.
The fabrication concludes with an inorganic removal of the hard mask materials.
The details of the fabrication can be found in the supplementary information (SI~\cite{supplementary}).

Our fabrication differs from earlier work \cite{khanaliloo_high-qv_2015, mitchell2019realizing, mouradian_rectangular_2017, wan_large-scale_2020, rugar_quantum_2021} in one main aspect: we demonstrate successful quasi-isotropic undercut of the waveguides at a considerably lower temperature of the reactor wafer table of only \qty{65}{\celsius}.
We show that the quasi-isotropic crystal-dependent reactive-ion etch in this temperature regime is successfully undercutting the waveguide structures without the need of an optional \ce{O2} anisotropic etch step following the vertical sidewalls coverage with \ce{Al2O3}. This has the key benefit of preserving the hard mask aspect ratio, without further edge mask rounding stemming from the \ce{O2} etch.

The fabricated devices consist of arrays of double-sided tapered waveguides, anchored to the surrounding bulk substrate by a square support structure, as seen in Fig.~\refsub{fig:fig1}{b}.
To couple light in and out of the waveguide we use optical fibers that are etched into conical tapers in hydrofluoric acid \cite{burek_fiber-coupled_2017}.
We position the fibers in front of the waveguide and exploit the lensing effect of the taper, as illustrated in Fig. \refsub{fig:fig1}{a}.
Though not as efficient as coupling through evanescent field, we choose this method as it allows easy variation of the distance between fiber and waveguide.
This will later be used to tune the phase of the reflected signal.
All experiments are performed at \qty{5}{\K} in a closed-cycle cryostat, with no external magnetic field. 
Within the SnV level structure (Fig. \refsub{fig:fig1}{c}), we focus on the optical zero-phonon line (ZPL) transition between the lower branches of the ground and excited states, of wavelength around $619$ nm.
Spontaneous emission from the excited state can also happen with a phonon assisted process, giving rise to a phonon sideband (PSB).

A photoluminescence excitation spectrum (PLE) (Fig.~\refsub{fig:fig1}{d}) reveals several SnV centers in the waveguide, which are spectrally resolvable owing to local variations of the strain environment.
We investigate the optical stability of one emitter (red arrow in Fig.~\refsub{fig:fig1}{d}) by performing consecutive PLE scans. 
The scans are pre-conditioned on a successful charge-resonance check~\cite{brevoord_heralded_2023}: Before each scan we turn on the probe laser at a set frequency and count how many PSB photons are detected. 
A threshold number of counts is chosen to make sure that the SnV is in the desired charge state and on resonance with the emitter (see SI~\cite{supplementary}). This heralding technique allows to mitigate effects of emitter ionization and of spectral diffusion~\cite{brevoord_heralded_2023}.

Summing data from $1.8$ hours of continuous measurement (Fig.~\refsub{fig:fig1}{e}), we observe an integrated linewidth of $(38.0\pm0.3)$~MHz, very close to the average linewidth of the single scans of $(32.1\pm0.1)$~MHz, indicating that there is very little effective spectral diffusion in our measurements. This can in principle be further improved by increasing the conditioning threshold at the expense of experiment speed. All the measurements reported below are conditioned on a charge-resonant check with similar threshold.
By measuring second-order photon correlations using different resonant laser powers (see SI~\cite{supplementary}), we extract the excited state lifetime of the emitter to be $(5.91\pm0.08)$~ns, corresponding to a transform-limited transition linewidth of $(26.7\pm0.3)$~MHz. This value is close to the average single scan linewidth, indicating that there is little residual broadening of the transition.

\begin{figure}[t]
    \centering
    \includegraphics[width = 0.4\textwidth]{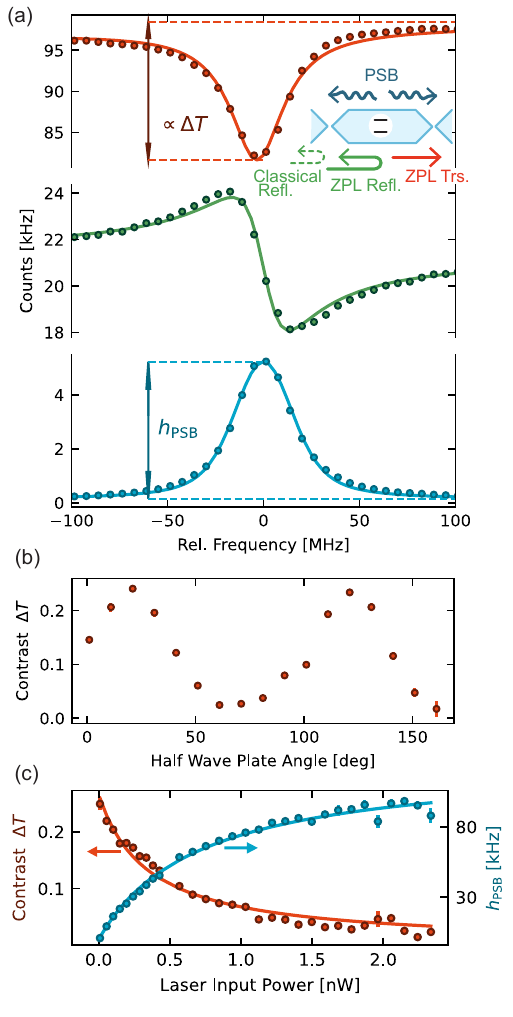}
    \caption{Spectroscopy of the waveguide-SnV system. (a) Simultaneous measurement of transmitted ZPL light through the system (top), reflected ZPL light (center) and PSB emission (bottom) while scanning the probe laser. The inset shows a schematic of the system, the colors represent the different channels measured. We highlight a relevant quantities used in the rest of the figure: the transmission contrast $\Delta T$ and the peak height of the PSB emission scan $h_\s{PSB}$. (b) Contrast of the transmission extinction when varying the input fiber polarization. (c) Transmission contrast and PSB peak height as a function of the power in the input fiber showing saturation of the SnV response when the average photon number increases. In the low power range, we observe a maximum transmission extinction of $0.25 \pm 0.01 $.}
    \label{fig:fig2}
\end{figure}

To probe the coupling of the SnV center to the waveguide, we scan the probe laser across the transition frequency, while simultaneously collecting both the transmitted and reflected signals. We spectrally filter the signals and record both ZPL and PSB (separately) in the reflected output port, and the ZPL in the transmission output port (see SI for details). This simultaneous measurement allows us to monitor the SnV behavior through the PSB emission, while observing its coherent interaction with the input probe through the ZPL signal (Fig. \refsub{fig:fig2}{a}).

We observe a significant extinction of the transmission signal on resonance, indicating a coherent light-matter interaction in our waveguide-QED system~\cite{shen_coherent_2005}: destructive interference between scattered photons and the transmitted field causes the emitter-induced reflection of single photons. 
The magnitude of the transmission dip contrast on resonance is determined by the emitter-waveguide coupling factor $\beta = \gamma_\s{wg}/\gamma_\s{tot}$, where $\gamma_\s{wg}\,(\gamma_\s{tot})$ is the decay rate into the waveguide (the total decay rate of the excited state). In particular, in the absence of dephasing of the optical transition, the transmission behavior is described by \cite{thyrrestrup_quantum_2018}
\begin{equation}
   T(\omega) = \left|1 - \frac{\beta}{ (1+\frac{\langle n \rangle}{n_c})(1+ \frac{2 i \omega}{\gamma_\s{tot}})} \right|^2,
    \label{eq:t_total}
\end{equation}

where $\omega$ is the detuning of the probe laser from the emitter, $\langle n \rangle$ is the average photon number per lifetime in the input state  and $n_\s{c} = \frac{1}{4\beta^2}$ is the critical photon number, which indicates saturation of the photon-emitter interaction.
In the limit of low excitation  ($\langle n \rangle \ll n_\s{c}$), the transmission contrast on resonance $\Delta T = 1-T(\omega=0)$ is thus related to the coupling factor $\beta$ as $\Delta T =  \beta(2-\beta)$.
Note that in this analysis, we ignore the small additional broadening of the optical transition due to dephasing, making our estimates for $\beta$ a strict lower bound. 

Experimentally, the value of the coupling factor $\beta$ can be reliably extracted by measuring the transmission contrast as a function of input laser power, given that $\langle n\rangle/n_c = P/P_c$ with $P_c$ the input power that saturates the interaction.  To ensure that we are optimally coupling the probe field with the linear dipole of the optical transition, we sweep the polarization of the input field to find the maximal transmission contrast (Fig. \refsub{fig:fig2}{b}). Fitting Eq. \eqref{eq:t_total} to the measured transmission contrast as a function of input power (Fig. \refsub{fig:fig2}{c}) we obtain $\beta=0.143\pm 0.005$.
This value is in good agreement with numerical simulations for our waveguide geometry (see SI~\cite{supplementary}) taking into account the emitter depth resulting from the implantation, a small lateral offset ($\approx 50$~nm) from the waveguide center, and the total efficiency of the transition of interest of $0.37$~\cite{herrmann_coherent_2023}, obtained combining quantum efficiency ($0.8$) \cite{iwasaki_tin-vacancy_2017}, Debye-Waller factor ($0.57$) \cite{gorlitz_spectroscopic_2020} and branching ratio between the two ZPL transitions ($0.8$) \cite{rugar_quantum_2021}.

The critical laser power at the fiber input, $P_\s{c}$, corresponds to the critical photon number $n_\s{c}$ at the SnV center: $P_\s{c} = \eta^{-1}\, h \nu n_\s{c} \gamma$ , where $\eta$ is the fiber-waveguide coupling efficiency, $\nu$ is the probe laser frequency and $\gamma$ is the decay rate related to the excited state lifetime. From the fit value 
$P_c = (0.32\pm 0.02)$~nW and knowing $n_\s{c}\sim12$ photons from the value of $\beta$, we determine the fiber-waveguide coupling efficiency to be $\eta=0.33\pm 0.02$.

\begin{figure}[t]
    \centering
    \includegraphics[width = 0.48\textwidth]{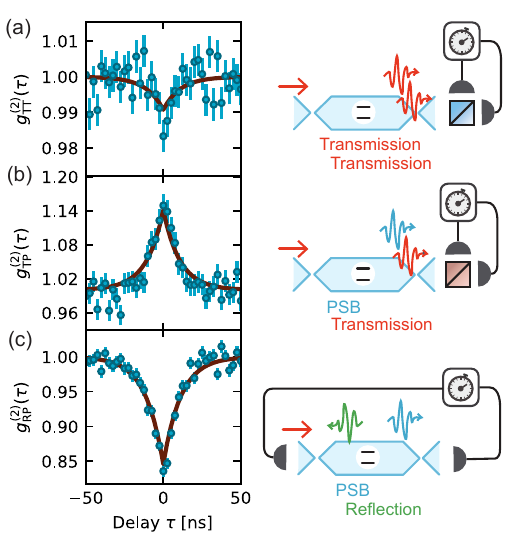}
    \caption{Second order correlation between (a) two transmitted photons, (b) a PSB and a transmitted photon, (c) a PSB and a reflected photon in the constructive interference regime (see main text and Fig.~\ref{fig:fig4} for details).
    The blue dots are experimental data and the red line is the theoretical model.}
    \label{fig:fig3}
\end{figure}

The reflection signal contains the single photons coherently reflected by the SnV center, interfering with classical reflection of the probe laser at the tapered fiber end.
In a simplified picture, considering a Lorentzian response of the SnV, the reflection signal can be modelled as \cite{bhaskar_quantum_2017, koch_super-poissonian_2022}
\begin{equation}
    R(\omega) = \left| 1 + \xi \frac{1}{1-2i\omega/\gamma_\s{tot}} e^{i\phi} \right|^2,
    \label{eq:reflection}
\end{equation}

\begin{figure}[h!]
    \centering
    \includegraphics[width = 0.48\textwidth]{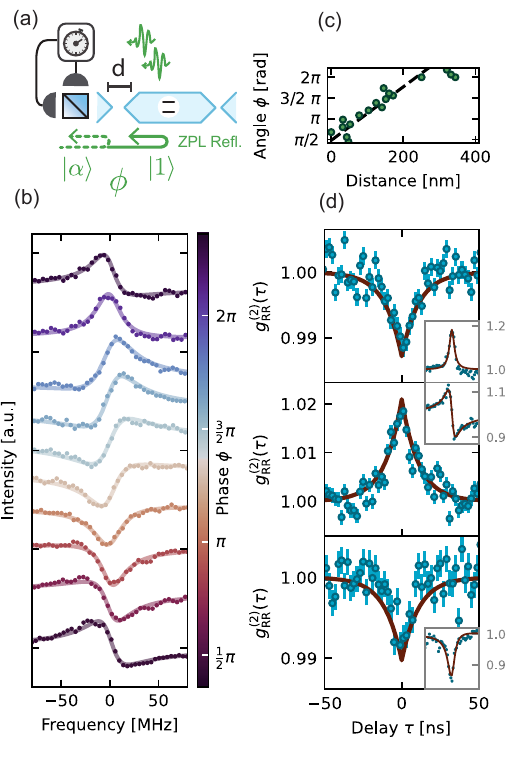}
    \caption{Reflection measurements. (a) Schematic of the measurement setting. (b) Reflection spectra at different fiber distance, corresponding to different relative phase between the single photon and coherent state components. The solid line is a fit with the simplified model of equation \ref{eq:reflection}, from this we extract the phase $\phi$ which is indicated by the colorbar. (c) Fitted phase as a function of the fiber displacement, measured by the positioner readout sensor. The black dashed line is a guide to the eye indicating a $2\pi$ phase change in the distance of half a wavelength. (d) Second order correlation of reflected photons in the constructive (top), dispersive (center) and destructive (bottom) interference regimes. The inset shows the reflection spectrum. We fit the reflection spectrum using Eq.~\eqref{eq:reflection} to extract $\phi$ and $\xi$, which we use as an input to the theoretical model. The red lines, both in the inset and the main figure, are the results of the theoretical model with the supplied parameters. The x axis in the inset is frequency, ranging from -80 MHz to 80 MHz.}
    \label{fig:fig4}
\end{figure}

where $\xi$ is the ratio  between the reflected single photons and the coherent state amplitude, which gives the average number of photons per lifetime in the input field, and $\phi$ is the phase difference between the coherent and single photon components.

The emitter-induced single-photon nonlinear reflection alters the photon statistics of the transmitted and reflected fields, as we expose below by measuring the second-order correlation for different combination of signals.

We start by correlating the transmitted signal with itself. The emitter can only reflect one photon per its optical lifetime. When two indistinguishable photons coherently scatter on the emitter within the timescale of its optical lifetime, the nonlinear interaction results in a strongly correlated two-photon bound-state that is perfectly transmitted \cite{shen_strongly_2007}. In the case of $\beta \to 1$ the wavefunction of the scattered light is dominated by this bound-state component, resulting in strong bunching of the transmitted light \cite{chang_single-photon_2007}. In our case however, as the decay rate in the waveguide is lower than the rate of other decay channels ($\beta < 0.5$), the bound-state contribution is suppressed compared to the uncorrelated scattering of the two photons involving other channels. In other words, the probability that one of the two photons is incoherently scattered is higher than the probability of emitting the two-photon bound state in the waveguide. In this regime, upon detection of a photon in the transmission signal the conditional probability of a second ZPL photon detection within the lifetime is suppressed compared to the steady state value. In full agreement with the predictions in Ref.~\cite{chang_single-photon_2007}, we find that this results in anti-bunching in the photon statistics for the transmission $g_\s{TT}^{(2)}$ (Fig. \refsub{fig:fig3}{a}).

Since we have access to one of the incoherent decay channels, namely the photons emitted in the PSB, we can verify that the effect on the photon statistics is induced by the coherent interaction of light with the emitter. We measure the probability of detecting a transmitted (Fig. \refsub{fig:fig3}{b}) or reflected photon (Fig. \refsub{fig:fig3}{c}) conditioned on the emission of a PSB photon. Detecting a PSB photon heralds an incoherent interaction which results in a higher probability of detecting a ZPL photon in the transmission port and a lower probability of detecting a ZPL photon in the reflected port, leading to the observed bunching and anti-bunching, respectively.

We compare the results in Fig.~\ref{fig:fig3} with numerical simulations (the red lines in the figures), where the system is modeled as a lossy cavity coupled to an emitter (see SI for details~\cite{supplementary}). Using values for the model parameters extracted from the data of Fig.~\ref{fig:fig2}, we find that the simulations accurately reproduce the behavior of the $g^{(2)}$ measurements.

Finally, we investigate the interference between the coherently scattered single photons and the reflected laser light in more detail.
By adjusting the fiber position relative to the waveguide, we are able to change the difference between the paths that the classical light and single photon components travel, and thereby controllably tune their relative phase $\phi$ (Fig. \refsub{fig:fig4}{a}). Figure \refsub{fig:fig4}{b} shows the variation in the reflection interference spectrum as we sweep the fiber distance, realized by applying a voltage on the piezo positioner.
The phase dependence on distance, which we extract from the piezopositioner sensor, is approximately linear and a full period is obtained in around half wavelength distance (Fig. \refsub{fig:fig4}{c}).

The photon statistics of the reflected signal depend on the relative phase $\phi$. 
We consider the three limit cases of constructive ($\phi = 6.28 \sim 2\pi$), dispersive ($\phi= 1.57 \sim\pi/2$) and destructive ($\phi= 3.41 \sim \pi$) interference.
In the constructive interference case, single photons are added to the coherent state, resulting in sub-poissonian photon statistics as evidenced by the measured anti-bunching in 
$g_\s{RR}^{(2)}$ in Fig. \refsub{fig:fig4}{d, top}.
In the dispersive and destructive interference cases, the presence of a non-zero phase makes the behavior of the photon statistics non-trivial as the relative phase is different for the one- and multi-photon components of the coherent state. Depending on the exact phase and the relative amplitudes of the single photons and the reflected coherent state, the relative weight of the one and two-photon components vary, resulting in either bunching or anti-bunching. In our regime we observe bunching for the dispersive interference Fig.~\refsub{fig:fig4}{d, middle} and weak antibunching for destructive interference Fig.~\refsub{fig:fig4}{d, bottom}. Numerical simulations (red lines) using our theoretical model show excellent agreement with the data.

In summary, we have presented a detailed investigation of a diamond SnV center coupled to a waveguide, showing significant transmission extinction, tunable interference between single photons and the reflected laser field, as well as providing insights into the nature of the emitter-induced changes in transmitted and reflected fields through photon correlation measurements. These results highlight diamond SnV centers integrated in waveguides as a promising platform for realising efficient integrated spin-photon interfaces.

Whereas nanophotonic cavities can provide overall much stronger interaction, the use of waveguides can alleviate significant fabrication overhead and by their broadband nature provide a more flexible platform, since they do not need to be tuned to the emitter frequency and readily allow for more centers to be used in the same device. We investigated four waveguides in this device and all contained suitably coupled SnV centers, with measured $\Delta T$ ranging between $15-34\%$.

While our work shows couplings that are in line with the state of the art for color center-waveguide systems \cite{bhaskar_quantum_2017, parker_diamond_2023, wan_large-scale_2020}, further improvement can be obtained by optimizing the emitter overlap with the optical mode: the waveguide thickness and the implantation depth can be matched to get the SnV closer to the center, while localized ion implantation could improve the lateral position.
Already at the established coupling, these devices, when combined with coherent spin control \cite{guo_microwave-based_2023, rosenthal_microwave_2023}, may allow for remote entanglement significantly surpassing the generation rates obtained the diamond NV center \cite{bernien_heralded_2013, hermans_quantum_2022}, opening up new avenues for scaling quantum networks.

\subsection{Acknowledgements}
We thank Henri Ervasti and Pieter Botma for software support, Kevin Chen for help in processing the Sn implanted sample, Anders S. Sørensen for helpful discussions and Yanik Herrmann and Julius Fischer for proofreading the manuscript.

We gratefully acknowledge support by the Dutch Research Council (NWO) through the Spinoza prize 2019 (project number SPI 63-264), by the Dutch Ministry of Economic Affairs and Climate Policy (EZK) as part of the Quantum Delta NL programme, by the joint research program “Modular quantum computers” by Fujitsu Limited and Delft University of Technology, co-funded by the Netherlands Enterprise Agency under project number PPS2007, and by the QIA-Phase 1 project through the European Union’s Horizon Europe research and innovation programme under grant agreement No. 101102140. L.D.S. acknowledges funding from the European Union’s Horizon 2020 research and innovation program under the Marie Sklodowska-Curie grant agreement No. 840393.

\textbf{Data availability:} The datasets that support this
manuscript are available at 4TU.ResearchData \cite{pasini_data_2023}.

\putbib
\end{bibunit}

\newpage

\begin{bibunit}

\title{SUPPLEMENTARY INFORMATION\\Nonlinear Quantum Photonics with a Tin-Vacancy Center Coupled to a One-Dimensional Diamond Waveguide}

\maketitle

\onecolumngrid

\justifying

\section{Sample and device fabrication}

\NewCommandCopy{\oldsection}{\section}
\renewcommand{\section}{\newpage\oldsection}

The sample fabrication process starts with pre-implantation surface treatment of a $<$100$>$ surface oriented electronic grade diamond substrate (Element 6). The sample substrate is first cleaned in a wet Piranha (ratio 3:1 of \ce{H2SO4 (95\%)}~:~\ce{H2O2 (31\%)}) inorganic solution for 20 min at \SI{80}{\celsius}, followed by the superficial $\sim$ \SI{5}{\um} etching via inductively-coupled-plasma reactive-ion-etching (ICP/RIE) \ce{Ar/Cl2} plasma chemistry based recipe in order to remove the residual polishing induced strain from the surface of the substrate. An additional $\sim$ \SI{5}{\um} ICP/RIE \ce{O2} chemistry based plasma etch is performed in order to remove residual chlorine contamination from the previous etching step \cite{ruf_cavity-enhanced_2021}. Following, the sample is inorganically cleaned in a Piranha solution (20 min at \SI{80}{\celsius}) and implanted with Sn ions (dose 1e11 ions/\ce{cm^{2}} with an energy of \SI{350}{keV}). Prior to the activation of the SnV centers by vacuum annealing (\SI{1200}{\celsius}), a triacid cleaning (ratio 1:1:1 of \ce{HClO4}$(70 \%)$~:~\ce{HNO3}$(70\%)$~:~\ce{H2SO4}$(>99\%)$) is performed for 1.5 hours in order to remove any residual organic contamination, followed by the the same wet inorganic cleaning procedure after the annealing step in order to remove any superficial graphite thin film layer formed during the annealing step of the diamond substrate. In order to assess the successful activation of the SnV centers, the sample is characterized prior to the nanofabrication of the suspended structures.

\begin{figure}[h]
    \centering
    \includegraphics[width=\linewidth]{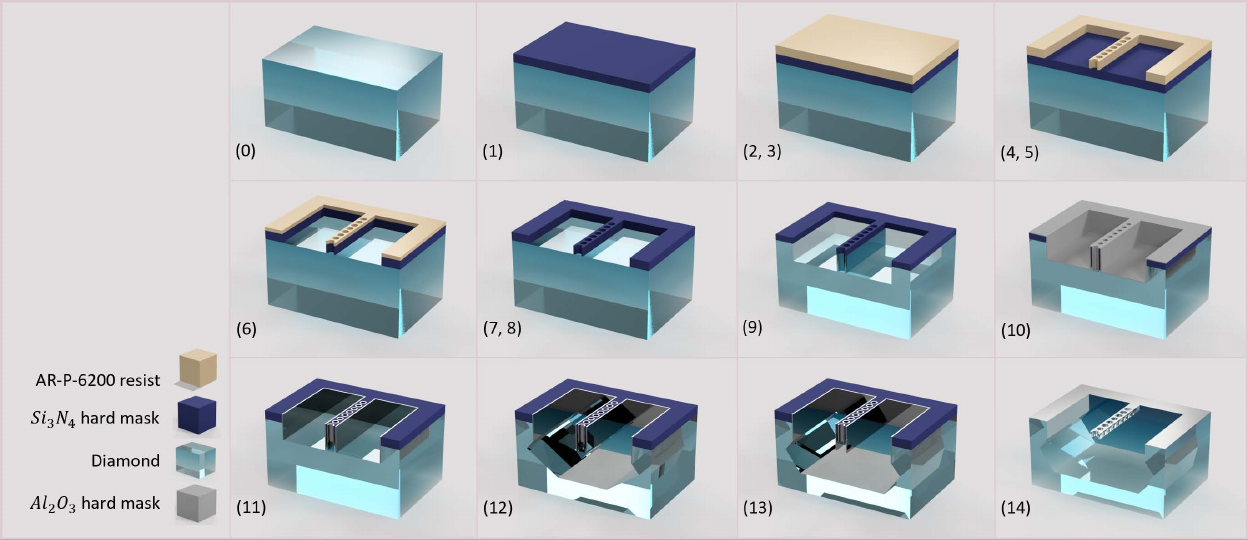}
    \caption{Nanophotonic structures fabrication process steps, based on Ref. \cite{khanaliloo_high-qv_2015, mitchell2019realizing, mouradian_rectangular_2017, wan_large-scale_2020, rugar_quantum_2021, ruf_cavity-enhanced_2021}. See text for details.}
    \label{fig:fabrication_steps}
\end{figure}

The nanofabrication of waveguiding structures follows the process based on the crystal-dependent quasi-isotropic-etch undercut method developed in references  \cite{khanaliloo_high-qv_2015, mitchell2019realizing, mouradian_rectangular_2017, wan_large-scale_2020, rugar_quantum_2021} and \cite{ruf_cavity-enhanced_2021}. A schematics of this is illustrated in Fig. \ref{fig:fabrication_steps}.

Specifically, we start with a pre-fabrication process extensive inorganic clean for 20 min in HF (40\%) at room temperature (Fig. \ref{fig:fabrication_steps}, step 0), followed by deposition and patterning of a hard mask material thin film layer of $\sim$ \SI{221}{nm} \ce{Si3N4} via plasma-enhanced-chemical-vapour-deposition (PECVD) (Fig. \ref{fig:fabrication_steps}, step 1). The waveguide structures design is longitudinally aligned with the  $<$110$>$ diamond crystallographic orientation and exposed via e-beam lithography of $\sim$ 450 nm of AR-P-6200.13 positive tone resist (Fig. \ref{fig:fabrication_steps}, steps 2 and 3). In order to avoid charging effects of the diamond substrate during the e-beam exposure, the surface of the e-beam resist is coated with $\sim$40 nm Electra 92 (AR-PC 5090) conductive polymer. Next, this is developed by immersing the sample in 60 s gentle stirring in \ce{H2O} and \ce{N2} blow-dry (to dissolve Electra 92), followed by 60 s DI water, 60 s pentyl acetate, 5 s ortho-xylene, 60 s isopropyl alcohol (IPA) (Fig. \ref{fig:fabrication_steps}, steps 4 and 5). The process proceeds with the transfer pattern into the \ce{Si3N4} hard mask material by means of ICP/RIE etch in a \ce{CHF3}/\ce{O2} based plasma chemistry (Fig. \ref{fig:fabrication_steps}, step 6) and complete removal of residual e-beam resist in a two-fold wet step: we first proceed with a resist removal in a PRS 3000 positive resist stripper solution, followed by extensive Piranha clean inorganic treatment for complete removal of organic material residues (Fig. \ref{fig:fabrication_steps}, steps 7 and 8). Such considerable inorganic cleaning is employed in order to prevent micro-masking within next dry etch steps that can potentially be caused by organic residues on the sample. 

The transfer pattern from the \ce{Si3N4} hard mask material is transferred by etching top-down in a dry ICP/RIE \ce{O2} plasma chemistry etch in the diamond substrate for an extent of $\sim$2.4x the designed thickness of the devices (Fig. \ref{fig:fabrication_steps}, step 9). It is crucial to note that the \ce{O2} dry etch heavily affects the aspect ratio of the patterned \ce{Si3N4} hard mask: we observe the etch rate of \ce{Si3N4} to be higher at the edges of the patterned structures compared to the determined etch rate of \ce{Si3N4} on flat area test samples, leading to enhanced erosion of the \ce{Si3N4} at the edges of the nanostructures. This yields rounded vertical sidewalls with few nanometers of hard mask material at the diamond to \ce{Si3N4} interface \cite{ruf_cavity-enhanced_2021}. This leads to weak points across the top surface of the hard mask that can compromise the integrity of the hard mask through the following fabrication steps. We circumvent this challenge by carefully tuning the trade-off between sufficient anisotropic diamond etch and \ce{Si3N4} mask integrity such that the last withstands the following \ce{O2} based dry etch steps foreseen by the overall fabrication process. 

Next, conformal atomic layer deposition (ALD) of $\sim$ 26 nm of \ce{Al2O3} for hard mask coverage of devices vertical sidewalls(Fig. \ref{fig:fabrication_steps}, step 10) and the horizontal coverage of \ce{Al2O3} is fully removed by ICP/RIE etch in a \ce{BCl3}/\ce{Cl2} plasma chemistry etch (Fig. \ref{fig:fabrication_steps}, step 11). This opens access to the diamond substrate: we directly follow with a two-step quasi-isotropic etch. For this, we employ the recipe developed in reference \cite{ruf_cavity-enhanced_2021}. In contrast to Ref. \cite{khanaliloo_high-qv_2015, mitchell2019realizing, mouradian_rectangular_2017, wan_large-scale_2020, rugar_quantum_2021}, here the quasi-isotropic etch plasma reactor wafer table temperature is only \SI{65}{\celsius}, therefore completing the full undercut and upward etch of the devices to the target thickness in 2 steps of 9 hours etch each (Fig. \ref{fig:fabrication_steps}, steps 12 and 13). The full release and upward quasi-isotropic total etch time is considerably long when compared to high temperature plasma regimes. On the other hand, here we demonstrate that this method successfully undercuts diamond waveguides in a low temperature regime. Finally, the hard mask materials are removed in an extensive inorganic treatment for 20 min in HF (40\%) at room temperature (Fig. \ref{fig:fabrication_steps}, step 14).

The overall qualitative characterization of the fabrication steps presented in this work has been executed on a scanning electron microscope Hitachi SEM Regulus system. The etch rates concerning the fabrication steps 6, 9 and 11 (Fig. \ref{fig:fabrication_steps}) have been pre-characterized on additional test samples, in parallel to the fabrication of the diamond sample employed in this work following similar methodology. Such etch tests have been conducted employing silicon substrate samples, with the thin films of interest deposited in parallel to the diamond substrate. Optical parameters and thickness of the employed materials have been determined via spectroscopic ellipsometry method on a Woollam M-2000 (XI-210) tool. The quasi-isotropic etch rate has been characterized employing supplementary diamond test samples (fabrication parallel to the sample in this work) and analyzed via SEM inspection.

\section{Experimental setup}
\begin{figure}[h]
    \centering
    \includegraphics[width=\textwidth]{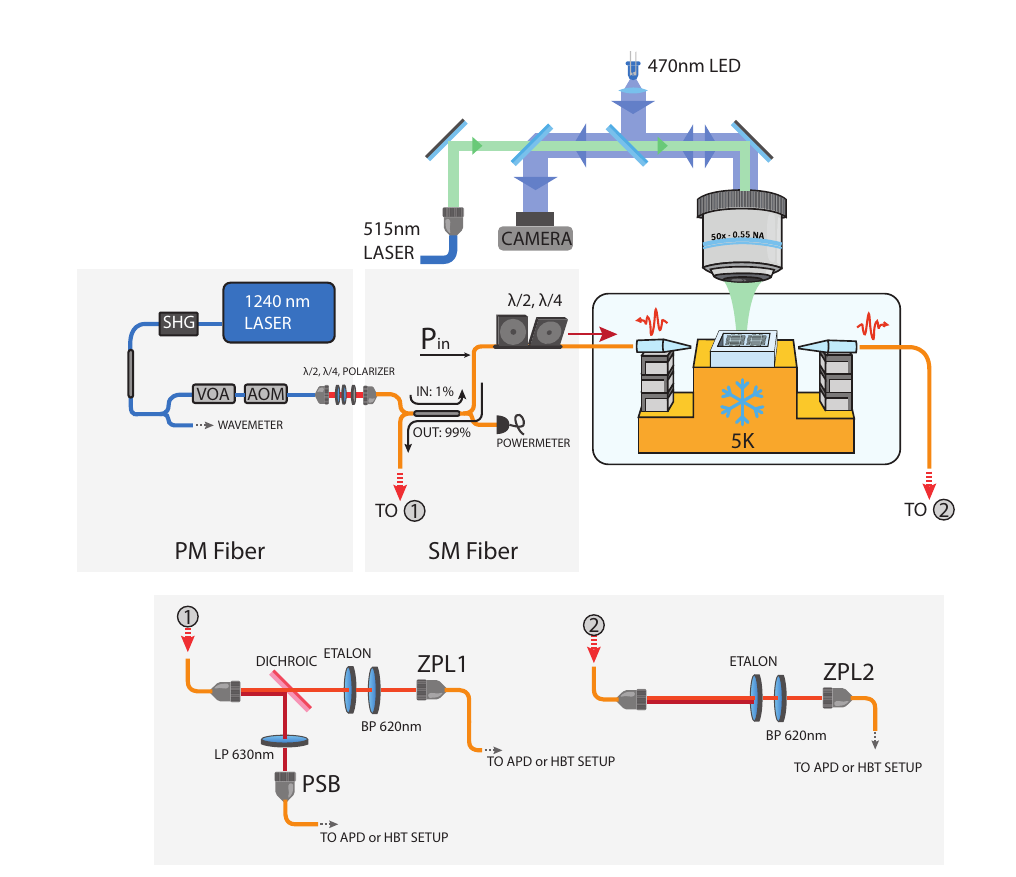}
    \caption{Experimental Setup. See text for explanation and abbreviations.}
    \label{fig:setup}
\end{figure}

Figure \ref{fig:setup} shows a schematic of our experimental setup. The sample is cooled to \SI{5}{K} in a closed cycle cryostat (Montana Instruments s50), together with two XYZ piezopositioner stacks (Attocube Systems ANPx51, ANPz51) which allow to control the position of the optical fibers.
We image the sample through an optical window in the cryostat using a long working distance objective (Mitutoyo 50X Plan Apochromat, NA: 0.55, WD: \SI{13}{mm}), which combined with an LED at $\sim$\SI{470}{nm} provides enough resolution to image the waveguides and the fiber tip. Through the same objective we deliver the repump laser pulses (Cobolt MLD-06 \SI{515}{nm}).
The resonant excitation and the photon collection is all via fiber. A tunable 1240 nm laser (Toptica DL Pro - TA) pumps a second harmonic generation fiber-coupled crystal (AdvR) to generate tunable laser light around \SI{620}{nm}. Part of the red output is sent to a wavemeter (High Finesse WS8) for frequency stabilization through a PID loop running in background and feeding back on the DL Pro controller to tune the IR frequency. The light used for resonant excitation goes through a manual Variable Optical Attenuator (VOA) and an Acousto Optical Modulator (AOM) used to generate the pulses. A short fiber-freespace-fiber module (Thorlabs FiberBench) with polarization control is used to bridge between polarization maintaining fiber of the laser control and the single mode fiber used in the rest of the experiment.
A 99:1 beam splitter is used to send $1\%$ of the laser light to the input fiber, where we use motorized polarization paddles to control the input polarization. The remaining $99\%$ is used for calibration and monitoring. The reflected light goes through the beam splitter and is collected in the $99\%$ return port.
We have two free-space setups for filtering and detection of the light. In the first, usually connected to the reflection port, a dichroic filter (Semrock FF625, short pass $\sim$ \SI{620}{nm}) separates the PSB and ZPL light. The PSB is further filtered by a Band Pass filter ($620\pm5$ nm) and a tunable narrow etalon filter (LightMachinery, $\sim45$ GHz FWHM) to get rid of the second ZPL transition. In the PSB path we further filter the excitation laser with two Long Pass tunable filters set approximately at \SI{630}{nm}. The second free-space stage only has the \SI{620}{nm} band pass and the etalon to filter around the ZPL.
The collected photons are fiber coupled and sent to avalanche photodiode (APD) single photon detectors (Lasercomponents COUNT) or to a Hanbury-Brown-Twiss (HBT) setup realized with a 50:50 fiber beam splitter and two APDs for correlation measurements.
We run the experiment using custom software based on the QMI \cite{raa_qmi_2023} package (version 0.41.0).

Our experiments use a lensed-fiber-like approach for coupling the waveguide to the tapered fiber. By optimising the position, we consistently measure a fiber-waveguide-fiber transmission efficiency up to $\sim4\%$, which translates to $\sim 20\%$ efficiency per side. Taking into account the potential losses at the crossing between the waveguide and the square support structure, this is in good agreement with our estimation of the excitation efficiency of the emitter of $\sim33\%$ (see main text) and hints to the fact that the emitter is located between the fiber taper and the support structure.

\section{Measurement sequences}

\begin{figure}[h]
    \centering
    \includegraphics[width=\linewidth]{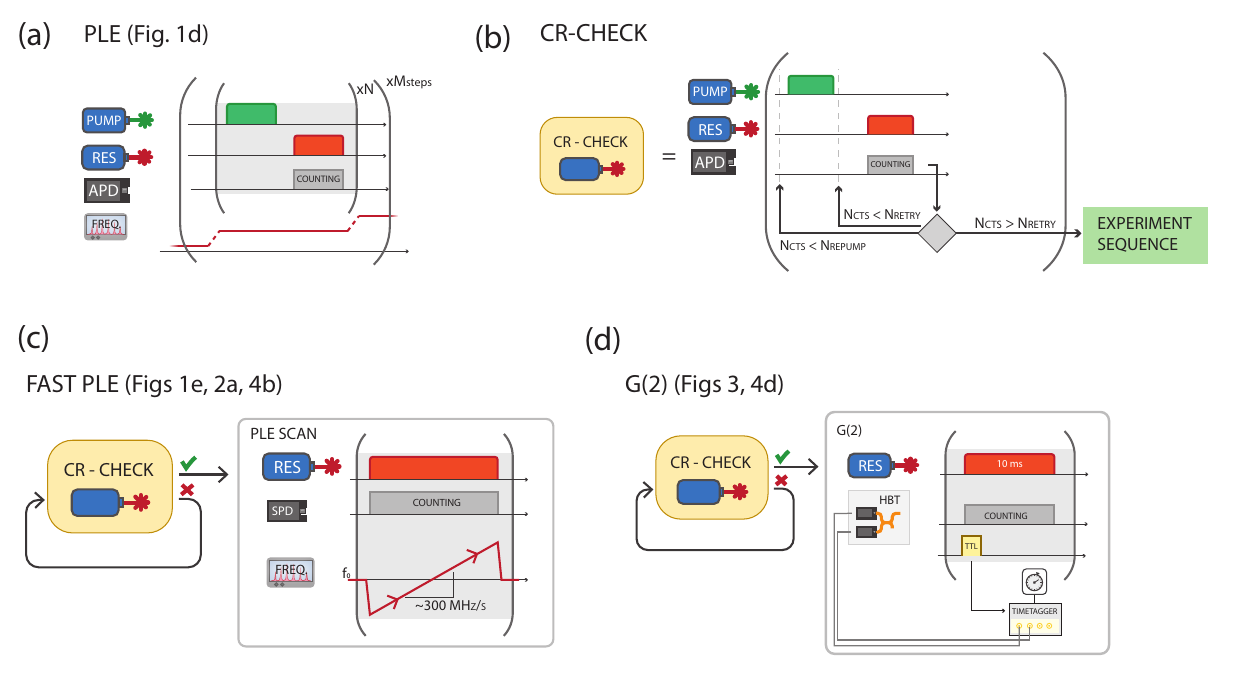}
    \caption{Measurement sequences}
    \label{fig:measurement_sequences}
\end{figure}

Figure \ref{fig:measurement_sequences} shows the measurement sequence for the experiments we realised. The control of the AOM pulses, fast laser frequency sweeps, APD counting, triggering of the time tagging hardware and logic operations are done by a Adwin Pro II microprocessor. 

The photoluminescence excitation (PLE) measurement in Fig. \refsub{fig:fig1}{d} is done with the sequence in \refsub{fig:measurement_sequences}{a}. 
We step the laser frequency and for each point we repeat a sequence consisting of a repump and a resonant pulse, during which we count the emitted photons.

To ensure that the SnV is in the correct charge and frequency state, we condition the measurements on a successful Charge-Resonance Check (CR-Check, Fig. \refsub{fig:measurement_sequences}{b}), see Ref~\cite{brevoord_heralded_2023} for details of the procedure.

We use CR-conditioned "fast" repeated PLE scans (Fig. \refsub{fig:measurement_sequences}{c}) to measure the PSB, transmission and reflection spectra in order to extract $\Delta T$ and the reflection parameters $\xi$ and $\phi$.
We apply an analog voltage to scan the laser frequency around a set point in a step-wise way.
In each step we integrate the counts of the APDs for $20$~ms.
The single scan speed is $\sim \qty{300}{\MHz/\second}$.

We perform the $g^{(2)}$ measurements with the sequence in Fig \refsub{fig:measurement_sequences}{d}, by repeatedly sending \qty{10}{\ms} resonant pulses, preceded by a trigger TTL pulse that serves as a sync signal for the timetagger.
The pulses are conditioned on a CR Check to ensure that the SnV is resonant with the probe.

\section{CR-Checked Linescans}
\begin{figure}[h]
    \centering
    \includegraphics[width=\linewidth]{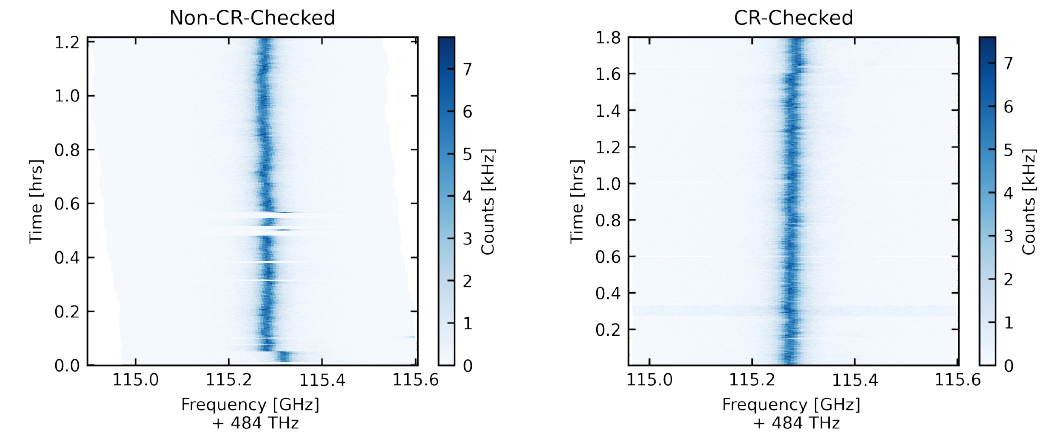}
    \caption{PLE measurement comparison. On the left, PLE scan where the presence or absence of a peak (at any frequency) in each scan conditions an off-resonant repump on the next scan. On the right, PLE scans that are pre-conditioned on a successful charge-resonance check.}
    \label{fig:linescans}
\end{figure}

Here we compare the PLE conditioned on Charge-Resonance (CR) check~\cite{brevoord_heralded_2023} (sequence in Fig. \ref{fig:measurement_sequences}(c), result on the right in Fig. \ref{fig:linescans}) with a PLE done with a less strict "brightness check" condition (left in Fig. \ref{fig:linescans}). In the latter, after each scan we check the maximum intensity of the PLE peak and compare it to the average countrate in the trace. If a threshold in the countrate is surpassed, indicating that there is a peak, we consider the emitter "bright" and we continue with the next scan. If the threshold is not met, we apply an off-resonant charge repump pulse (\SI{515}{nm}) before the next scan.
In the trace on the left in Fig. \ref{fig:linescans} one can notice some frequency jumps and empty scans. The "brightness check" only probabilistically brings the emitter in the right charge state, but that can also be at a different frequency since the off-resonant pulse can modify the charge environment around the SnV. As long as the emitter is bright, in this case, it does not show significant spectral diffusion. Note that the condition only acts on the following scan.
The CR-Check, instead, is a pre-condition. The scan only starts once we know that the emitter is in the right charge and resonance state. Therefore, each trace shows a peak at the target frequency. Conditioning the experiments with this technique prevents the recording of data when the emitter is detuned due to a frequency jump or absent due to ionization.
For the Non-CR-Checked linescans we did not lock the laser frequency after every scan.
In contrast to the CR-Check linescans, the frequency of the probe Laser drifts over time.
This is visible as a drift of the center frequency of the scan-range.
The frequency reported in the scan is measured at the wavemeter, that we constantly monitor, showing that the emitter frequency is stable.

\section{Lifetime measurement from resonant $g^{(2)}_\s{PP}(\tau)$}

\begin{figure}[h]
    \centering
    \includegraphics[width=\linewidth]{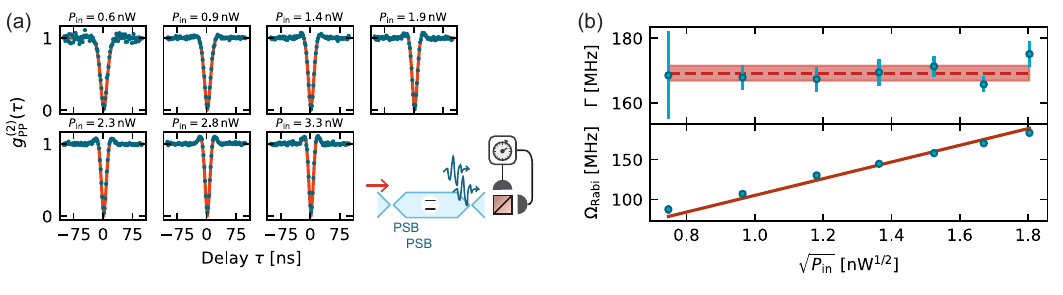}
    \caption{
    (a)
    Second order correlation between two PSB photons measured at various powers. The light blue curve is a fit using equation \ref{eq:resonant_g2}
    (b)
    From the fit in (a) we extract the decay rate (top) and Rabi frequency (bottom) as a function of the probe laser power. We obtain the decay rate by a weighted average of the values (dashed red line in the top graph, shadowed area indicates the uncertainty). The red line in the bottom graph is a linear fit.
    }
    \label{fig:lifetimeg2}
\end{figure}

To verify that the optical transition is transform limited, we extract the optical decay rate from a series of resonant second order correlation measurements at different powers.
We resonantly excite the transition with $10$ ms laser pulses and measure the second order correlation of the emitted PSB photons $g_\s{PP}^{(2)}$ (Fig. \refsub{fig:lifetimeg2}{a}).
We vary the power of the excitation pulse and for each measurement and fit the data with the function \cite[p. 208]{steck_quantum_2007}
\begin{equation}
    g_\s{PP}^{(2)}(\tau)
    =1-e^{-(3 \Gamma / 4) \tau}\left(\cos \Omega_{\Gamma} \tau+\frac{3 \Gamma}{4 \Omega_{\Gamma}} \sin \Omega_{\Gamma} \tau\right)
    + C
    \label{eq:resonant_g2}
\end{equation}.

At the lowest measured power, $g_\s{PP}^{(2)}(0) = 0.036 \pm 0.019$, showing that we are measuring a single emitter.
Additionally, we extract the optical Rabi frequency and the decay rate from the fits. The top graph in Figure \refsub{fig:lifetimeg2}{b} shows the decay rate for each measurement.
We take the weighted average, including the uncertainty, to extract a decay rate $\Gamma$ of ($168\pm2$)~MHz, which corresponds to a excited state lifetime of ($5.91\pm0.08$)~ns and to a transform-limited transition linewidth of ($26.7\pm0.3$)~MHz.
This matches well with our average single scan linewidth, indicating that there is little residual broadening of the transition.
The bottom plot shows the Rabi frequency, which scales linearly as the square root of the probe power.
This measurement is also performed with  CR-checks in between excitation pulses.

\section{Input Polarization}

\begin{figure}[h]
    \centering
    \includegraphics[width=\linewidth]{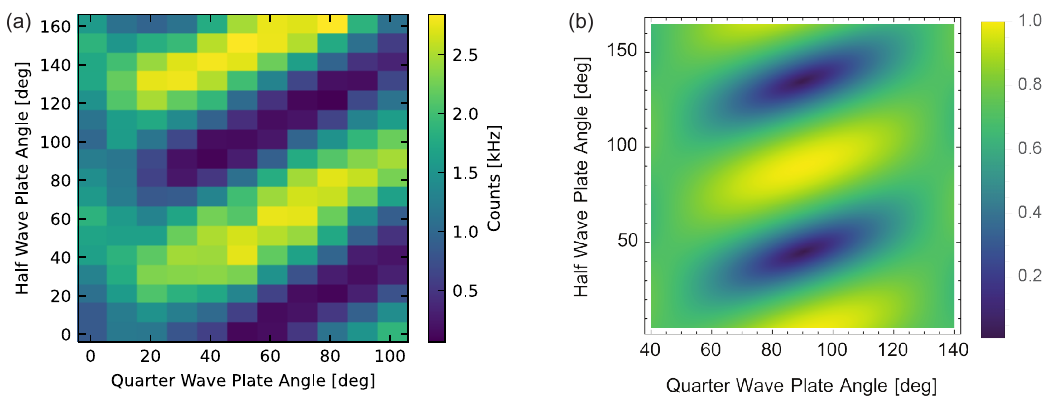}
    \caption{
    Effect of the probe laser polarization on detected PSB counts.
    (a)
    Varying the angles of a quarter and half wave plate in the path of the probe laser changes the amount of detected PSB counts.
    (b)
    A model of the effect based on Jones calculus.
    }
    \label{fig:polarization_sweeps}
\end{figure}

Here we vary the polarization of the probe laser by changing the angles of two fiber paddles acting as a $\lambda/4$ and $\lambda/2$ waveplate, respectively (see Fig. \ref{fig:setup}). 
At each combination of angles, we record 30 PLE scans and fit the summed PSB counts with a Lorentzian, as done in Fig. \refsub{fig:fig2}{a}.
In Fig. \refsub{fig:polarization_sweeps}{a} plot the height of that Lorentzian $h_\mathrm{PSB}$ (we filter out the failed fits, mostly happening in the area where the counts are low).
When the polarization of the probe laser is aligned with the dipole moment of the SnV, the PSB counts are maximal.
Rotating the half wave plate by $\sim$\ang{45} from that optimal setting, makes the polarization of the probe laser orthogonal to the dipole moment of the SnV, this is were the counts are minimal.
We model this behaviour using the Jones calculus \cite{collett_field_2005}.
Here the waveplates are described by 2x2 matrices

\begin{equation}
    M_{\lambda/2}(\phi) = 
    e^{- i \pi /2}
    \begin{pmatrix}
        \cos^2\phi - \sin^2\phi & 
        2 \sin\phi \cos\phi \\
        2 \sin\phi \cos\phi &
        -\cos^2\phi + \sin^2\phi
    \end{pmatrix}
\end{equation}
and
\begin{equation}
    M_{\lambda/4}(\phi) = 
    e^{- i \pi /4}
    \begin{pmatrix}
        \cos^2\phi + i\sin^2\phi & 
        (1-i) \sin\phi \cos\phi \\
        (1-i) \sin\phi \cos\phi &
        i \cos^2\phi + \sin^2\phi
    \end{pmatrix},
\end{equation}

where $\phi$ is the rotation angle of the waveplate.
We act with both waveplates on vertical polarization and then project on vertical polarization.

\begin{align*}
    h_\s{PSB}^\s{model}
    (\phi_1, \phi_2)
    &= 
    \| 
    \begin{pmatrix}
        1 \\ 0
    \end{pmatrix}
    ^\s T
    M_{\lambda/2}(\phi_1) \,
    M_{\lambda/4}(\phi_2)
    \begin{pmatrix}
        1 \\ 0
    \end{pmatrix}
    \|\\
    &=
    \frac{1}{2} \sqrt{\cos \left[4 \, (\phi_1-\phi_2)\right]
    +\cos (4 \phi_2)+2}
\end{align*}

The modeled behaviour is shown in Fig.~\refsub{fig:polarization_sweeps}{b} and reproduces well the measured data, except for an absolute shift in the waveplates angles. Note that since we don't know the absolute polarization at the location of the fiber paddles the plot ranges of experiment and model are not the same.
Also the fiber paddles do not induce perfect $\lambda/2$ and $\lambda/4$ retardation. 

\section{Simulations of SnV-Waveguide coupling}

We compare the measured $\beta$ factor with finite element simulations in COMSOL Multiphysics.
The waveguide is assumed to be rectangular of width \SI{300}{nm} and height \SI{250}{nm}, which is the approximate dimension of the devices for this sample measured from SEM images.
The propagation direction of the waveguides is along the $\langle100\rangle$ crystallographic direction of the diamond.
This means that half of the SnVs will have the transition dipole on the plane perpendicular to the waveguide and with an angle of \ang{54.25} from the vertical axis.
We then simulate the SnV as a point electric dipole with the correct orientation.
The coupling $\beta$ can be obtained as \cite{thyrrestrup_quantum_2018}

\begin{equation}
    \beta = \frac{\Im[E_{\hat{d}} (x_0, y_0, z\to\infty)]}{\Im[E_{\hat{d}} (x_0, y_0, z_0)]}
\end{equation}

where $E_{\hat{d}}$ is the electric field in the direction of the dipole and $(x_0, y_0, z_0)$ is the position of the dipole.

The field at the dipole location is obtained from the simulation. 
To obtain the field at $z\to\infty$ we make the waveguide $\sim10$ micrometers long and analyze $E_{\hat{d}}$ in the last few micrometers of propagation.
We fit the electric field with a sum of sinusoidal functions to account for the coupling to different modes allowed in the waveguide (ideally one TE and one TM mode).
The predominant mode is considered to be the TE mode, we extract the $\beta$ for this mode.
We do this for different positions of the dipole in the waveguide: at the center, where the maximum coupling is expected, and at the implantation depth of \SI{88}{nm} with three different offsets from the center in the x direction ($\Delta x$) of 0, 50 and \SI{100}{nm}.
For all of these we report the coupling for an ideal dipole $\beta_\text{ideal}$. 

When the system is not an ideal two level transition, the effective coupling is decreased by other possible decay channels from the excited state, which reduce the maximum possible decay rate in the waveguide $\gamma_\text{wg}$.
In our case, we consider three decay channels other than the optical transition of interest: radiative compared to non-radiative decays (Quantum Efficiency, QE), emission in the ZPL compared to PSB (Debye-Waller factor, DW), and decay in the ZPL transition of interest compared to the one from lower branch of the excited state to upper branch of the ground state (Branching Ratio, BR). We use an efficiency $\zeta = \text{QE}\times\text{DW}\times\text{BR} = 0.37$ \cite{herrmann_coherent_2023}, from which we calculate $\beta_\text{eff} = \frac{\zeta \gamma_\text{wg}}{\gamma_\text{tot}} =  \zeta \beta_\text{ideal}$.

We obtain the following values:

\begin{table}[h]
\begin{tabular}{|c|c||c|c|}
    Depth [nm] &  $\Delta x$ [nm] & $\beta_\text{ideal}$ & $\beta_\text{eff}$\\
    \hline
     125 & 0 & 0.606 & 0.224 \\
     88 & 0 & 0.507 & 0.187 \\
     88 & 50 & 0.403 & 0.149 \\
     88 & 100 & 0.256 & 0.095 \\
\end{tabular}
\end{table}

We observe that the measured coupling for our emitter, $\beta=0.143$ agrees very well with the SnV being located at the implantation depth and offset from the center of the waveguide by $\sim$\SI{50}{nm}.

\section{Theoretical models}

\subsection{Analytical model for transmission and reflection fitting}

For the analytical formula describing the behavior of the transmission signal and used for the fits we follow closely the theory in \cite{javadi_single-photon_2015, thyrrestrup_quantum_2018}. 
The transition linewidth of our SnV is very close to the transform limited value expected from the lifetime measurement, so we choose to neglect dephasing in the transmission dip analysis. 
In the presence of dephasing, the transmission extinction contrast for a fixed $\beta$ is decreased. 
For a dephasing rate $\gamma_\s{deph}$ this scales as
\begin{equation}
    \Delta T = \frac{\beta(2-\beta)}{1+2\frac{\gamma_\s{deph}}{\gamma_\s{tot}}}.
\end{equation}
This makes our $\beta$ a lower bound, since if dephasing is present we would need a stronger coupling to explain the contrast we measure.
For the reflection spectrum we assume a simple single mode interference with a lorentzian response of the emitter, similar to what observed in \cite{bhaskar_quantum_2017, koch_super-poissonian_2022}.

\subsection{Model for the numerical simulations of the $g^{(2)}$}

\begin{figure}[h]
    \centering
    \includegraphics{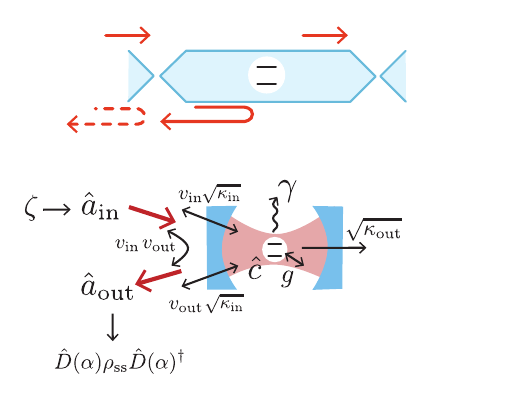}
    \caption{Schematic of the theoretical model. Above a sketch of the input and output channels of the waveguide-SnV system, below the corresponding modes and coupling parameters.}
    \label{fig:theorymodel}
\end{figure}

For numerically simulating the $g^{(2)}$, we model the SnV-waveguide system as a lossy cavity coupled to a two level system as depicted in Fig. \ref{fig:theorymodel}. 
The transmission signal can be obtained by the decay of the cavity mode $\hat{c}$ in the output of the cavity. 
For modeling the reflection we use a similar approach as in \cite{kiilerich_input-output_2019}, where two modes $\hat{a}_\textrm{in}$ and $\hat{a}_\textrm{out}$ couple to the cavity. 
We add a coherent drive on mode $\hat{a}_\textrm{in}$ to model the laser field. 
Part of the cavity mode which does not interact with the emitter can leak out in the reflection mode $\hat{a}_\textrm{out}$. 
This artifact coming from the cavity approximation is unwanted as in the waveguide the non-interacting light should go through. 
We tune the cross-coupling term $v_\textrm{in}v_\textrm{out}$ to cancel out the unwanted leaking light from the cavity.
$\hat{a}_\textrm{out}$ now represents the ideal reflection from the waveguide-QED system.
The dynamics of the system is described by the Hamiltonian:

\begin{align}
    H &= H_0 + H_\s{drive} + H_\s{JC} + H_\s{three-mode}\\
    H_0 &= (\omega_e - \omega) |e\rangle\langle e| + (\omega_c - \omega)\left[ \hat{c}^\dag \hat{c} + \hat{a}_\s{in}^\dag \hat{a}_\s{in} + \hat{a}_\s{out}^\dag \hat{a}_\s{out} \right]\\
    H_\s{drive} &= i\zeta (\hat{a}_\s{in}^\dag - \hat{a}_\s{in})\\
    H_\s{JC} &= ig(\hat{c} |e\rangle\langle g| + h.c.)\\
    H_\s{three-mode} &= (\sqrt{\kappa_\s{in}}v_\s{in} \hat{a}_\s{in}^\dag \hat{c} +
    \sqrt{\kappa_\s{in}}v_\s{out} \hat{c}^\dag\hat{a}_\s{out} +
    v_\s{in}v_\s{out} \hat{a}_\s{in}^\dag\hat{a}_\s{out} + h.c.)
\end{align}

and the jump operators:

\begin{align}
    L_\s{emitter} &= \sqrt{\gamma}|e\rangle\langle e|\\
    L_\s{cavity} &= \sqrt{\kappa_\s{in} + \kappa_\s{out}} \hat{c}\\
    L_\s{three-mode} &= \sqrt{\kappa_\s{in}} \hat{c} + v_\s{in} \hat{a}_\s{in} +  v_\s{out} \hat{a}_\s{out}.
\end{align}

The "lossy cavity" approximation of the waveguide is realised by having all coupling terms between the optical modes larger than the emitter decay rate, $(\kappa_\s{in}+\kappa_\s{out}), v_\s{in}, v_\s{out} \gg \gamma$. Since we are interested in the low excitation regime, we truncate the Hilbert space to two excitations.
We match the measured waveguide coupling with the weak-cavity parameters, we calculate the effective cooperativity corresponding to $\beta$:
\begin{equation}
    C = \frac{\beta}{1-\beta} = \frac{4 g^2}{\kappa \gamma}.
\end{equation}\\
We use the QuTiP Python package \cite{johansson_qutip_2012} to numerically solve the master equation for the density matrix and the steady state of the system $\rho_{ss}$.\\
To model the interference between the scattered photons and the classical light from the input laser, we use a displaced single photon state \cite{de_oliveira_properties_1990}.
We displace the steady state density matrix with a displacement operator $\hat{D}(\alpha)$, the complex amplitude $\alpha$ has a phase $\phi$ which encodes the phase difference between the single photon and coherent components, and amplitude $|\alpha|$ that is set in order to match the single-photon/coherent state ratio $\xi$ as measured experimentally.
We transform the Hamiltonian and jump operator to work on a displaced frame:
\begin{align}
    H & \to \hat D(\alpha)H \hat D(\alpha)^\dag\\
    L & \to \hat D(\alpha)L \hat D(\alpha)^\dag.
\end{align}
The operators we use for transmission, refection and PSB are:
\begin{align}
    \hat{t} &= \sqrt{\kappa_\s{out}} \hat{c}\\
    \hat{r} &= \hat D(\alpha)\hat{a}_\s{out} \hat D(\alpha)\\
    \hat{\sigma}_e &= |e\rangle\langle e|.
\end{align}

For every simulation, we first reproduce the spectrum by feeding the model with the measured parameters $\beta, \phi, \xi$ and measuring the expectation value for $\hat{t}^\dag\hat{t}, \hat{r}^\dag\hat{r}$ and $\hat{\sigma}_e$. Then, we calculate the second order correlation between two arbitrary operators $\hat{O}_1, \hat{O}_2$ as:
\begin{equation}
    g^{(2)}_{\hat{O}_1\hat{O}_2}(\tau) = \frac{ \s{Tr} \{ \hat{O}_2^\dag\hat{O}_2 e^{\tau \mathcal{L}} (\hat{O}_1\rho_{ss}\hat{O}_1^\dag) \} } { \s{Tr}
    \{ \hat{O}_1\rho_{ss}\hat{O}_1^\dag \} \s{Tr} 
    \{\hat{O}_2\rho_{ss}\hat{O}_2^\dag\} }
\end{equation}
Where $\mathcal{L}$ is the Liouvillian superoperator calculated in the displaced frame.

\section{Author contributions}

M.P. and T.T. performed the experiments and analyzed the data. N.C. designed and fabricated the device. N.C. and M.P. fabricated the tapered fibers. M.P. designed the setup and built it with C.P. . 
A.R.M., T.T., C.P. and M.P. characterized the device. M.P. developed the theoretical model and simulations with input from J.B.. L.D.S., H.K.C.B., J.M.B. and C.W. contributed to the development of the measurement infrastructure. M.P., T.T. and R.H. wrote the manuscript with input from N.C.. All authors commented on the manuscript. R.H. supervised the experiment.

\putbib
\end{bibunit}

\end{document}